\pdfoutput=1
\documentclass[10pt,twocolumn,english,sort,twocolumn]{IEEEtran}
\usepackage{helvet}

\usepackage[T1]{fontenc}
\usepackage[latin9]{inputenc}
\usepackage[letterpaper]{geometry}
\usepackage{color}
\usepackage{babel}
\usepackage{units}
\usepackage{graphicx}
\usepackage{esint}
\usepackage[unicode=true,pdfusetitle,
 bookmarks=true,bookmarksnumbered=false,bookmarksopen=false,
 breaklinks=false,pdfborder={0 0 0},backref=false,colorlinks=false]
 {hyperref}

\makeatletter
\renewcommand{\fnum@figure}{Fig.~\thefigure}

\makeatother

\begin{document}

\title{Invertibility of spectral x-ray data with pileup--two dimension-two
spectrum case}

\author{Robert E. Alvarez\thanks{\protect\href{http://www.aprendtech.com}{aprendtech.com}, ralvarez@aprendtech.com}}
\maketitle
\begin{abstract}
In the Alvarez-Macovski\cite{Alvarez1976} method, the line integrals
of the x-ray basis set coefficients are computed from measurements
with multiple spectra. An important question is whether the transformation
from measurements to line integrals is invertible. This paper presents
a proof that for a system with two spectra and a photon counting detector,
pileup does not affect the invertibility of the system. If the system
is invertible with no pileup, it will remain invertible with pileup
although the reduced Jacobian may lead to increased noise.

\vspace{0.2cm}

\hspace{-0.1cm}Key Words: Jacobian, photon counting, dead time, pileup,
spectral x-ray, dual energy, energy selective
\end{abstract}

\section{Introduction}

In the Alvarez-Macovski\cite{Alvarez1976} method, the line integrals
of the x-ray basis set coefficients are computed from measurements
with multiple spectra. The introduction of photon counting detectors
into medical x-ray imaging\cite{taguchi2013vision} gives the possibility
of providing the spectral data by pulse height analysis (PHA). These
detectors, however, have multiple defects\cite{Overdick2008} that
affect the information they provide. In some cases, the defects have
been found to produce sharply increased noise in the estimates of
the line integrals at specific object attenuation\cite{alvarez2017noninvertibility}.
The increased noise may indicate potential non-invertibility of the
transformation between the spectral measurements and the line integrals.
Therefore, it is important to develop mathematical descriptions of
the invertibility of the transformation.

This paper is a step towards this mathematical description. It presents
a proof that measurements with two spectra and a photon counting detector
with pileup do not affect the invertibility of the system. If the
system is invertible with no pileup, it will remain invertible with
pileup although the reduced Jacobian may lead to increased noise.
An example of the system analyzed is measurements with a photon counting
detector of the transmitted flux with two different x-ray tube voltages.
Note that the results of this paper are not applicable to pulse height
analysis (PHA) since pileup changes the effective spectra of the bins\cite{AlvarezSNRwithPileup2014}.
My recent paper\cite{alvarez2017noninvertibility} gives an example
of a three bin PHA system that becomes non-invertible for high pileup.

This paper addresses only the invertibility with deterministic, non-noisy
measurements although it describes the conditioning of the system,
which also affects noise.

\section{The Alvarez-Macovski method\label{sub:Alv-Mac-method}}

For biological materials we can approximate the x-ray attenuation
coefficient $\mu(\mathbf{r},E)$ accurately with a two function basis
set\cite{alvarez2013dimensionality} 
\begin{equation}
\mu(\mathbf{r},E)=a_{1}(\mathbf{r})f_{1}(E)+a_{2}(\mathbf{r})f_{2}(E).\label{eq:2-func-decomp}
\end{equation}

\noindent{}In this equation, $a_{i}(\mathbf{r})$ are the basis set
coefficients and $f_{i}(E)$ are the basis functions, $i=1\ldots2$.
As implied by the notation, the coefficients $a_{i}(\mathbf{r})$
are functions only of the position $\mathbf{r}$ within the object
and the functions $f_{i}(E)$ depend only on the x-ray energy $E$.
If there is a high atomic number contrast agent, we need to extend
the basis set to higher dimensions. 

Neglecting scatter, the expected value of the number of transmitted
photons $\lambda_{k}$ with an effective measurement spectrum $S_{k}(E)$
is 
\begin{equation}
\lambda_{k}=\int S_{k}(E)e^{-\int_{\mathcal{L}}\mu\left(\mathbf{r},E\right)d\mathbf{r}}dE\label{eq:Ik-integral}
\end{equation}

\noindent{}where the line integral in the exponent is on a line $\mathcal{L}$
from the x-ray source to the detector. 

Using the decomposition, Eq. \ref{eq:2-func-decomp}, the line integral
in Eq. \ref{eq:Ik-integral} is 
\begin{equation}
\int_{\mathcal{L}}\mu\left(\mathbf{r},E\right)d\mathbf{r}=A_{1}f_{1}(E)+A_{2}f_{2}(E).\label{eq:L(E)-A1-f1-A2-f2}
\end{equation}

\noindent{}where $A_{i}=\int a_{i}\left(\mathbf{r}\right)d\mathbf{r},\ i=1\ldots2$
are the line integrals of the basis set coefficients. Summarizing
the $A_{i}$ as the components of the A-vector, $\mathbf{A}$, and
the basis functions at energy $E$ as a vector $\mathbf{f}(E)=\left[f_{1}(E),\ f_{2}(E)\right]$,
we can write the line integral as the inner product of $\mathbf{A}$
and $\mathbf{f}(E)$, 
\begin{equation}
\int_{\mathcal{L}}\mu\left(\mathbf{r},E\right)d\mathbf{r}=\mathbf{A\bullet f}(E).\label{eq:Line-integral-dot-prod}
\end{equation}
The measurements are summarized by a vector, $\mathbf{N}$, whose
components are the expected photon counts with each effective spectrum.
Since the body transmission is exponential in $\mathbf{A}$, we can
approximately linearize the measurements by taking logarithms. The
results is the log measurement vector $\mathbf{L}=-\log(\mathbf{N/\mathbf{N_{0}}})$,
where $\mathbf{N_{0}}$ is the expected value of the measurements
with no object in the beam and the division means that corresponding
members of the vectors are divided. 

Equations \ref{eq:Ik-integral} define a relationship between $\mathbf{A}$
and the expected value measurement vector, $\mathbf{L(A)}$. The invertibility
of this transformation is the subject of this paper.

\section{Invertibility without pileup--simple cases\label{sec:Invertibility-2dim-2spect-no-pileup}}

Before discussing the invertibility in general, two simple but important
cases will be discussed.

\subsection{Delta function spectra}

The first case is two monoenergetic spectra with energies $E_{1}$
and $E_{2}$. In this case, $L_{i}=\mathbf{A\bullet f}(E_{i}),\ i=1,2$
and the transformation can be written $\mathbf{L(A)=MA}$ where $\mathbf{M}$
is a matrix with coefficients
\[
\mathbf{M}=\left[\begin{array}{ccc}
f_{1}(E_{1}) &  & f_{2}(E_{1})\\
f_{1}(E_{2}) &  & f_{2}(E_{2})
\end{array}\right]
\]

\noindent The transformation is invertible if the determinant of
$\mathbf{M}$ is not equal to zero: 
\begin{equation}
\left|{\begin{array}{cc}
{f_{1}(E_{1})}\hfill & {f_{2}(E_{1})}\hfill\\
{f_{1}(E_{2})}\hfill & {f_{2}(E_{2})}\hfill
\end{array}}\right|\neq0
\end{equation}

\noindent That is if
\begin{equation}
\frac{f_{1}(E_{1})}{f_{2}(E_{1})}\neq\frac{f_{1}(E_{2})}{f_{2}(E_{2})}\label{eq:f ratios}
\end{equation}

\begin{figure}
\noindent{}\includegraphics[scale=0.8]{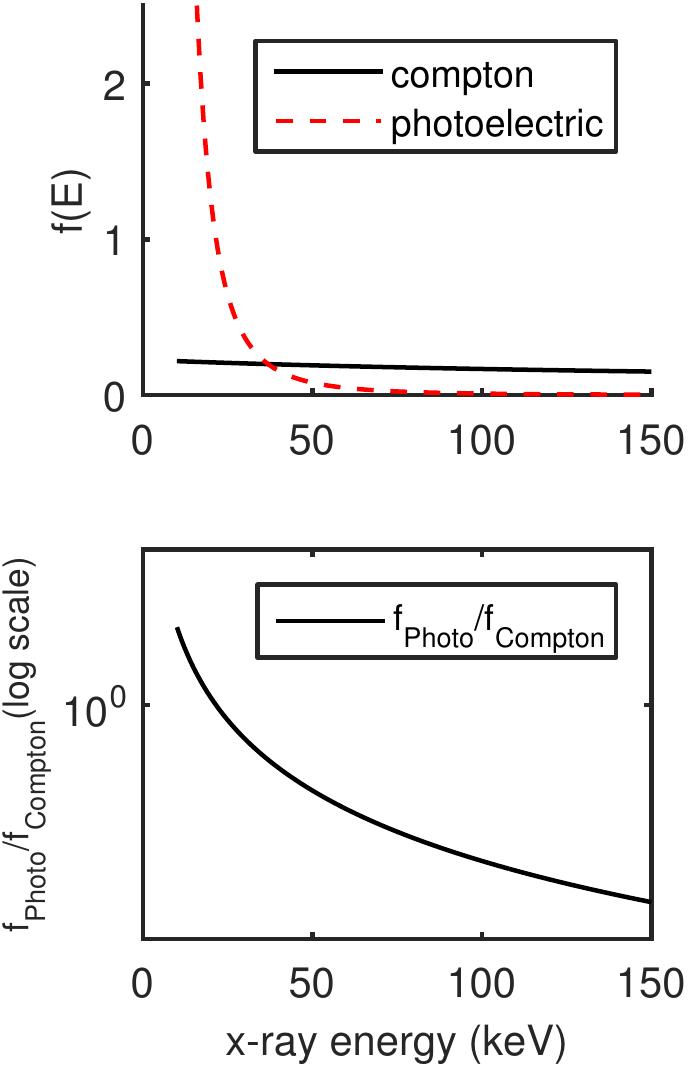}

\protect\caption{Ratio of the Compton scattering/photoelectric cross-sections basis
set functions (bottom graph). Since the ratio is monotonically decreasing,
the ratios in Eq. \ref{eq:f ratios} are different if the delta function
energies are different.\label{fig:Ratio-of-optimal funcs}}
\end{figure}

\noindent The bottom graph of Fig. \ref{fig:Ratio-of-optimal funcs}
shows the ratio of the Compton scattering/photoelectric cross-sections
basis set functions\cite{Alvarez1976} versus energy in the medical
diagnostic region. Note that the ratio is monotonically decreasing.
Thus, the condition in equation (\ref{eq:f ratios}) will be true
if the two energies are different. The $\mathbf{M}$ matrix for any
other basis set that spans the space of attenuation coefficients will
be $\mathbf{M'=TM}$ where $\mathbf{T}$ is an invertible matrix so
it has a nonzero determinant. Since these are square matrices, $\det\left(\mathbf{M'}\right)=\det\left(\mathbf{T}\right)\det\left(\mathbf{M}\right)$
and if $\det\left(\mathbf{M}\right)\neq0$ then $\det\left(\mathbf{M'}\right)\neq0$
so this result is true for any valid basis set.

\subsection{Single material object\label{sub:Single-material-object}}

Another useful special case is when the object is known to consist
of a single material with basis set coefficient vector $\mathbf{a}$.
In this case, a single photon count measurement suffices to determine
the material thickness. With a single material of thickness $t$,
the $\mathbf{A}$ vector is 
\[
\mathbf{A=a}t
\]

\noindent{}and the expected number of photons is a function of only
$t$
\[
\lambda(t)=\int S_{1}(E)exp[-\mu(E)t]dE.
\]

\noindent{}The derivative of $\lambda$ with respect to $t$ is always
negative
\begin{equation}
\begin{array}{ccccc}
\left\langle \lambda\right\rangle ' & = & \frac{d\lambda}{dt}\\
 & = & -\int\mu(E)S_{1}(E)exp[-\mu(E)t]dE<0.
\end{array}\label{eq:lambda-prime}
\end{equation}

\noindent{}because the integrand is always greater than zero. Since
$\lambda(t)$ is monotonically decreasing, it can be inverted to compute
$t$.

\section{Invertibility for the general case with no pileup}

\begin{figure}
\centering{}\includegraphics[scale=0.6]{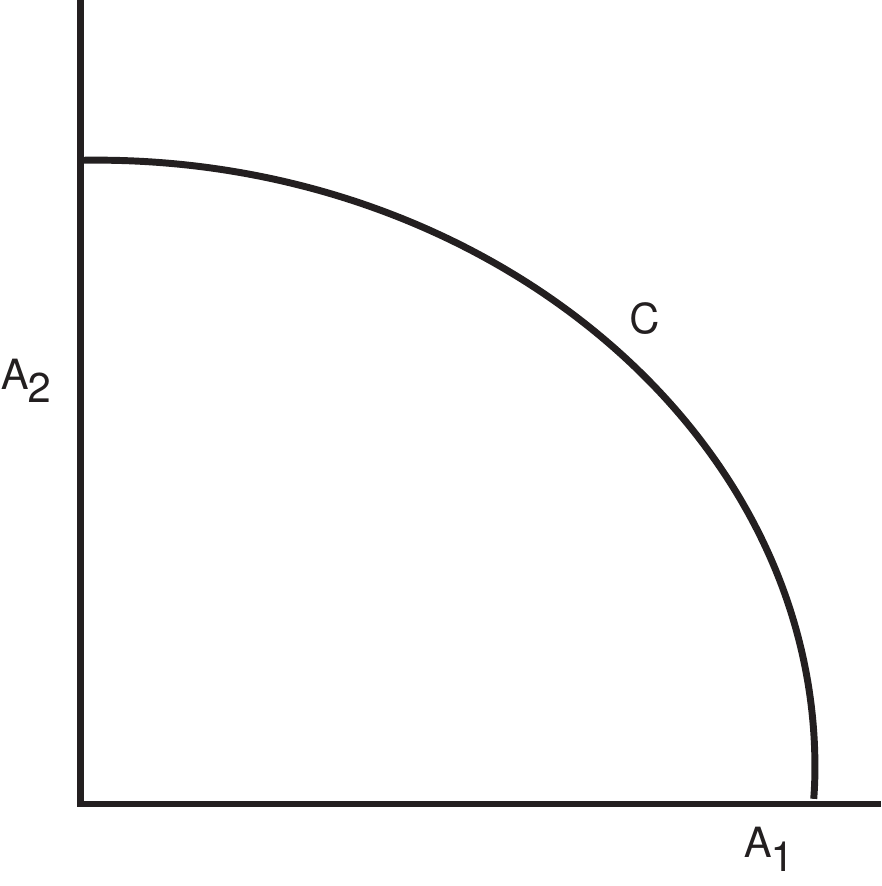}

\protect\caption{Closed contour used in proof of invertibility. The contour consists
of segments of the two positive axes and a quadrant of a circle joining
the ends of the line segments. \label{fig:Closed-contour-4InvertProof}}
\end{figure}

For the general case, the following theorem is useful (Fulks 1978\cite{Fulks1978}
page 284): 
\begin{quotation}
Let F be a continuously differentiable mapping defined on an open
region D in E2, with range R in E2 , and lets its Jacobian be never
zero in D. Suppose further that C is a simple closed curve that, together
with its interior (recall the Jordan curve theorem), lies in D, and
that F is one-to-one on C. Then the image T of C is a simple closed
curve that, together with its interior, lies in R. Furthermore, F
is one-to-one on the closed region consisting of C and its interior,
so that the inverse transformation can be defined on the closed region
consisting of T and its interior.
\end{quotation}
\noindent{}which I paraphrase as
\begin{quotation}
If the Jacobian of a continuously differentiable two dimensional mapping
is nonzero throughout an open region D and if the mapping is one to
one on a simple closed curve C which lies in D, then the mapping is
one to one on C and its interior. 
\end{quotation}
The first quadrant will be used as the region with the closed curve
$C$ consisting of segments along the positive axes and a circle joining
the ends of the segments, as shown in Figure \ref{fig:Closed-contour-4InvertProof}.
This is a region of theoretical and practical importance because a
basis set consisting of the attenuation coefficients of the calibration
materials is usually used, Since only positive equivalent thicknesses
of the calibration materials can be used, this region must contain
all the measured values. . 

In the theorem, the Jacobian is the determinant of the matrix of all
the partial derivatives of the transformation. Since $\mathbf{L}$
is the logarithm of the measurements, the components of its Jacobian
matrix are
\begin{equation}
\begin{array}{ccccc}
M_{ij} & = & \frac{\partial L_{i}}{\partial A_{j}}\\
 & = & -\frac{\int f_{j}(E)S_{i}(E)e^{-\mathbf{A\bullet f}(E)}dE}{\int S_{i}(E)e^{-\mathbf{A\bullet f}(E)}dE}\quad i,j=1,2
\end{array}\label{eq:Mij}
\end{equation}

\noindent Note that by defining the normalized spectra 
\begin{equation}
\hat{s}_{i}(E)=\frac{S_{i}(E)e^{-\mathbf{A\bullet f}(E)}}{\int S_{i}(E)e^{-\mathbf{A\bullet f}(E)}dE}\quad i=1,2
\end{equation}

\noindent the $M_{ij}$ are 
\[
M_{ij}=\int f_{j}(E)\hat{s}_{i}(E)e^{-\mathbf{A\bullet f}(E)}dE
\]

\noindent{}That is, they are the effective values of the basis functions
in the spectra transmitted through the object.

\subsection{Application of the theorem to the dual energy transformation}

To apply the first part of the theorem, the transformation must be
shown to be invertible on a closed curve in the domain. The simple
cases discussed Sec. \ref{sec:Invertibility-2dim-2spect-no-pileup}
may be used for this proof. 

The parts of the curve along the axes are special cases of the single
material case. Each axis corresponds to different thicknesses of one
of the basis materials if the attenuation coefficient of the calibration
material is used as a basis function. If other basis functions are
used, the coordinates can be transformed to the attenuation coefficients
of real materials and the proof applies in the transformed coordinates. 

The circle of large radius is an approximation of the single energy
case. For large radius, there will be high attenuation. With beam
hardening, the transmitted spectrum with large attenuation therefore
approaches the two monoenergetic spectra case where the energies are
the maximum energies in the spectra. If the maximum energies are different,
we can approach the known invertible monoenergetic case arbitrarily
closely by making the radius larger and larger. 

The remaining part of the theorem requires us to show that $J\left(\mathbf{A}\right)$
is non-zero inside $C$. This must be tested with individual spectra.
The following sections show that, for the system studied, pileup does
not affect this condition. If the Jacobian is non-zero without pileup,
it will also be non-zero with pileup.

\section{Expected number of photons recorded with pileup}

\textcolor{black}{The response time of a photon counting detector
is modeled using the dead time, $\tau$, which is defined to be the
minimum time between two photons that are recorded as separate events\cite{Knoll2000}.
The dead time is an abstraction that combines the contributions to
the response time of all the physical effects in the detector. In
this model, the detector is assumed to start in a ``live'' state.
With the arrival of a photon, the detector enters a separate state
where it does not count additional photons. The non-paralyzable model
 will be used where the time in the separate state is assumed to be
fixed and independent of the arrival of any other photons during the
dead time. There is a second model commonly used, called paralyzable,
where the arrival of photons extends the time in the non-counting
state. Both models give similar recorded counts at low interaction
rates but give different results at high rates where the probability
of multiple interactions during the dead time becomes significant.
Measurements by Taguchi et al.\cite{taguchi_modeling_2011} indicate
that for the detectors they studied the non-paralyzable model is more
accurate at higher count rates. It also leads to simpler analytical
results\cite{yu_fessler_PMB_2000}.}

In a previous paper\cite{AlvarezSNRwithPileup2014}, I used the central
limit theorem of renewal processes\cite{ross_stochastic_1995} to
show that, for recording times much greater than the mean inter-event
time, the expected value of recorded counts with pileup approaches
\begin{equation}
\left\langle N_{rec}\right\rangle =\frac{\lambda}{1+\eta}.\label{eq:Mean-w-deadtime}
\end{equation}

In this equation, $\lambda$ is the expected number of photons incident
on the detector during the measurement time and $\eta$ is the expected
number of photons arriving during the dead time $\tau$. If $\rho$
is the average rate of photon arrivals then $\eta=\rho\tau$. If $T_{rec}$
is the measurement time then $\rho=\nicefrac{\lambda}{T_{rec}}$.
Defining $b=\nicefrac{\tau}{T_{rec}}$, $\eta=b\lambda$ and the expected
recorded counts are 
\begin{equation}
\left\langle N_{rec}\right\rangle =\frac{\lambda}{1+b\lambda}.\label{eq:Nrec}
\end{equation}

\section{Invertibility of two basis functions-two spectra case with pileup}

In this section I give a proof that with photon counting detector
measurements of the total number of transmitted photons with two spectra,
pileup does not affect invertibility. An example would be making two
sequential measurements of an object using an x-ray tube with different
voltages. Note that this does not prove that any two spectrum measurement
with pileup is invertible. For example with two bin photon counting
with PHA, pileup causes the recorded spectrum to change so the assumptions
of this section would not be met.

The proof is analogous to the proof for the measurements with no pileup
described in the previous Sec. \ref{sec:Invertibility-2dim-2spect-no-pileup}.

\subsection{Invertibility with pileup on the contour $C$}

First, I will show that if the transformation is invertible on the
path in the first quadrant shown in Fig. \ref{fig:Closed-contour-4InvertProof}
without pileup it is also invertible with pileup.

The proof for invertibility on the circular segment joining the segments
on the axes is also applicable with pileup since, for large thicknesses,
the count rate is very low so the pileup parameter $\eta$ is essentially
equal to zero and the pileup counts are the same as those without
pileup. 

Next we need to show that the data are invertible on the paths from
the origin along the coordinates axes. The equation for the expected
value of the recorded number of photons with pileup is 
\begin{equation}
\left\langle N_{rec}\right\rangle =\frac{\lambda}{1+b\lambda}.\label{eq:Nrec-w-pileup}
\end{equation}
Differentiating this equation along the axes
\[
\left\langle N_{rec}\right\rangle '=\frac{\lambda'}{1+b\lambda}-\frac{b\left(\lambda'\right)\lambda}{\left(1+b\lambda\right)^{2}}
\]

\noindent{}where a prime denotes a derivative with respect to object
thickness. Factoring the equation
\begin{equation}
\begin{array}{ccc}
\left\langle N_{rec}\right\rangle ' & = & \frac{\lambda'}{1+b\lambda}\left(1-\frac{b\lambda}{1+b\lambda}\right)\\
 & = & \frac{\lambda'}{\left(1+b\lambda\right)^{2}}
\end{array}.\label{eq:Nrec-prime}
\end{equation}

\noindent{}From Eq. \ref{eq:lambda-prime}, for a single material
$\lambda'$ is always less than zero. Since the denominator of Eq.
\ref{eq:Nrec-prime} is always positive, the derivative of the recorded
counts with pileup with is never equal to zero so the transformation
of the recorded counts with pileup is invertible along the coordinate
axes.

This shows that if the transformation without pileup is invertible
on $C$ then the transformation with pileup is also invertible on
the curve.

\subsection{Jacobian inside $C$}

The log measurement with pileup is 
\[
L_{rec,i}=-\log\left(\frac{N_{rec,i}}{N_{rec,i,0}}\right)
\]

\noindent{}and the Jacobian matrix with pileup has elements
\begin{equation}
\begin{array}{ccc}
M_{rec,ij} & = & \frac{\partial L_{rec,i}}{\partial A_{j}}\\
 & = & -\frac{1}{N_{rec,i}}\frac{\partial N_{rec,i}}{\partial A_{j}}.
\end{array}\label{eq:Mrec-i}
\end{equation}

\noindent{}From Eq. \ref{eq:Nrec}
\[
N_{rec,i}=\frac{\lambda_{i}}{1+b\lambda_{i}}
\]

\noindent{} and from Eq. \ref{eq:Nrec-prime}
\[
\frac{\partial N_{rec,i}}{\partial A_{j}}=\frac{\lambda_{i}'}{\left(1+b\lambda_{i}\right)^{2}}.
\]

Substituting in Eq. \ref{eq:Mrec-i}
\begin{equation}
M_{rec,ij}=\frac{1}{1+b\lambda_{i}}\left[\frac{\lambda'_{i}}{\lambda_{i}}\right]\label{eq:Mrec-ij}
\end{equation}

\noindent{}From the definition in Eq. \ref{eq:Mij}, the term in
brackets in Eq. \ref{eq:Mrec-ij} is the element of the Jacobian matrix
without pileup. Therefore, 
\[
M_{rec,ij}=\frac{M_{0,ij}}{1+b\lambda_{i}}
\]

\noindent{}Since for the dual energy case the matrices are $2\times2$,
the determinant is 
\[
\begin{array}{ccc}
J_{rec} & = & M_{rec,11}M_{rec,22}-M_{rec,12}M_{rec,21}\\
 & = & \frac{J_{0}}{\left(1+b\lambda_{1}\right)\left(1+b\lambda_{2}\right)}
\end{array}.
\]

\noindent{}where $J_{0}$ is the Jacobian without pileup. Since the
terms in the denominator are always positive, the zero values of Jacobian
determinant with pileup, if any, will occur at the same points as
the Jacobian without pileup.

\section{Discussion}

The data acquisition model used is unrealistic---most systems with
photon counting detectors would also use PHA. Nevertheless, it provides
an example of invertibility with pileup. 

As discussed in my previous paper\cite{AlvarezSNRwithPileup2014},
due to the exponential probability distribution of x-ray photon inter-arrival
times, pulse pileup is a fundamental effect. It will always be present
in photon counting systems no matter how small the response time.
Pileup has accurate analytical models and its analysis may lead to
an understanding of invertibility with the other detector defects
such as charge trapping and sharing, polarization, and incomplete
photon energy deposition due to Compton scattering and K radiation
escape\cite{Overdick2008}.

\section{Conclusion}

A proof is given that pileup does not affect the invertibility of
a two spectrum, photon counting data with pileup system.


\end{document}